# Anomalous Optoelectronic Properties of Chiral Carbon Nanorings…and One Ring to Rule Them All

Bryan M. Wong* and Jonathan W. Lee

Materials Chemistry Department, Sandia National Laboratories, Livermore, California 94551

*Corresponding author. E-mail: bmwong@sandia.gov. Web: http://alum.mit.edu/www/usagi

Carbon nanorings are hoop-shaped, $\pi$-conjugated macrocycles which form the fundamental annular segments of single-walled carbon nanotubes (SWNTs). In a very recent report, the structures of chiral carbon nanorings (which may serve as chemical templates for synthesizing chiral nanotubes) were experimentally synthesized and characterized for the first time. Here, in our communication, we show that the *excited-state* properties of these unique chiral nanorings exhibit anomalous and extremely interesting optoelectronic properties, with excitation energies growing larger as a function of size (in contradiction with typical quantum confinement effects). While the first electronic excitation in armchair nanorings is forbidden with a weak oscillator strength, we find that the same excitation in chiral nanorings is allowed due to a strong geometric symmetry breaking. Most importantly, among all the possible nanorings synthesized in this fashion, we show that *only one ring*, corresponding to a SWNT with chiral indices ($n+3,n+1$), is extremely special with large photoinduced transitions that are most readily observable in spectroscopic experiments.

**Table of Contents Graphic:**

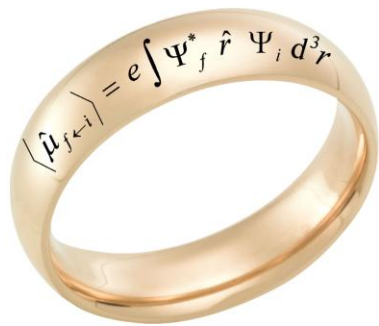

Organic nanostructures consisting of delocalized $p$ orbitals are one of the most studied classes of photovoltaic materials due to a unique electronic structure which can support mobile charge carriers throughout the highly-conjugated $\pi$-electron system. Although linear polymer chains[1] and quasi-1D nanostructures[2,3] have shown great utility in optoelectronic devices, cyclic organic materials are of particular interest due to their conserved and symmetrical structures. For example, recent synthetic advances in the creation of various cyclic oligothiophenes have opened up the possibility of forming self-assembled nanostructures[4,5] with specific sizes and chemical functional groups. Similarly, there has also been a significant effort to synthesize various carbon nanorings since they form the fundamental annular segments of carbon nanotubes and may further serve as templates for synthesizing nanotubes with specific chiralities.[6,7] Motivated by this line of reasoning, the Bertozzi group recently synthesized and characterized the very first series of cycloparaphenylenes,[8] which are composed solely of phenyl rings sequentially connected to form a single "nanohoop." Their synthesis was especially noteworthy since cycloparaphenylenes represent the smallest sidewall segment of armchair ($n,n$) single-walled carbon nanotubes (SWNTs). In a series of related studies, we[9] and the Sundholm group[10,11] theoretically characterized the optoelectronic properties of these nanorings and surprisingly found that the lowest excitation energy grows larger as a function of nanoring size, in contradiction with typical confinement effects. This unusual behavior in nanorings is even more surprising given the fact that the optical spectra of carbon nanotubes *do* obey expected quantum confinement trends with excitation energies decreasing with nanotube diameter. To explain this anomalous effect in the nanoring systems, both of our previous studies unanimously concluded that the high strain in the smaller cycloparaphenylenes is taken at the expense of aromaticity/stability in the individual phenyl rings; as a result, the optical gap of the smaller nanorings is narrowed relative to the (less-strained) larger nanorings. Unfortunately, as discussed in our previous report,[9] the direct experimental observation of this lowest excitation energy is extremely difficult since all of these excitations are optically-inactive "dark" states (the transition is optically forbidden since both the ground and first excited state have the same electronic parity), due to the highly-symmetric geometries of the cycloparaphenylenes.

Very recently, the Itami group carried out the first experimental characterization of a chiral nanoring, synthesized by effectively inserting an acene subunit within the cycloparaphenylene backbone[12] (cf. Figure 1). These researchers further showed that by inserting various acenes with appropriate linkages, *all* possible chiral structures of nanorings can be created. As can be easily seen in Figure 1, the chemical insertion of an acene unit effectively serves as a molecular "kink" in the nanoring and completely breaks the symmetry of the original cycloparaphenylene. As a result, one can intuitively expect that the anomalous optical trends (which were optically dark) in the symmetric cycloparaphenylenes can now be more readily observed in these asymmetric chiral nanorings.

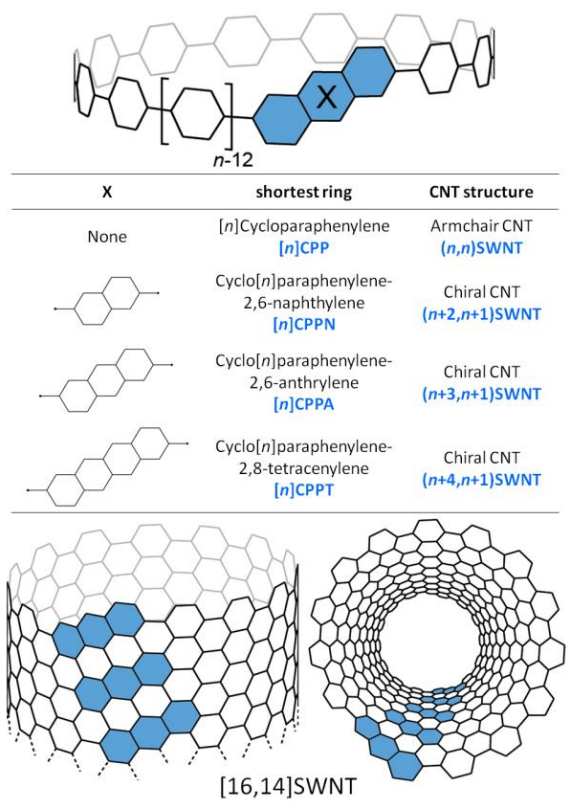

**Figure 1.** Geometric structures of chiral carbon nanorings. The ($n,m$) chirality indices of the full SWNTs which correspond to the individual nanorings are listed next to their chemical names.

Here we present a detailed study of the excited-state properties of these chiral nanorings, focusing on the subtle electronic effects which allow the lowest excited-state transition to be spectroscopically observed. In this study, we find that both the geometric symmetry-breaking of the chiral nanorings combined with the relative alignment of orbital energies in these structures strongly influences which of these nanorings have the most intense transitions.

We first describe the systems considered in our calculations which are illustrated and categorized in Figure 1. Using the same terminology as the Itami group,[12] cycloparaphenylenes composed of $n$ phenyl rings are denoted by [$n$]CPP. Chiral nanorings which are formed by inserting either a naphthalene, anthracene, or tetracene unit are denoted by [$n$]CPPN, [$n$]CPPA, and [$n$]CPPT, respectively. In the last column of Figure 1, the chirality indices of the full SWNTs which correspond to the individual nanorings are also listed. It is important to note, however, that while we list the ($n,m$) chirality indices in Figure 1 for easy comparison to the "parent" carbon nanotube structure, we strongly emphasize that the anomalous trends discussed in this work are only unique to these novel nanoring structures and should not (and can not) be extrapolated to the optoelectronic properties for the parent carbon nanotube.

All quantum chemical calculations in this study utilized both density functional theory (DFT) and time-dependent DFT (TD-DFT) calculations in conjunction with the B3LYP hybrid functional. Previous investigations by us[9,13] and Tretiak *et al.*[14,15] have shown that the B3LYP kernel provides a balanced description of neutral excitons in these systems and in other organic polymers. We also explored other functionals, including new range-separated methods we have extensively tuned for organic photovoltaics,[16-18] and found that the anomalous optoelectronic trends in our study were not affected by our specific choice of functional (we will return to this point during our later discussion on alignment of orbital energies in these nanostructures). All ground-state geometries were re-optimized at the B3LYP/6-31G(d,p) level of theory using the initial geometries from Refs. 12 and 19, and harmonic frequencies at the same level of theory were calculated to verify these stationary points were local minima. Strain energies of the nanorings were also calculated using homodesmotic reactions,[12] and these results are tabulated in the Supporting Information. At each optimized geometry, a time-dependent DFT calculation was carried out at the B3LYP/6-31G(d,p) level of theory to obtain the lowest singlet vertical excitation energy.

Table 1 lists the lowest excitation energies and oscillator strengths for the various nanorings, and Figure 2 displays these overall trends as a function of size. As Figure 2a clearly shows, the lowest excitation energy for all of the nanorings increases with the number of benzene rings. Again, this unusual trend is due to an electronic competition between maintaining the aromaticity of the individual benzene rings versus minimizing the strain energy along the backbone chain.[9,10] In all of the smaller nanorings, the highly-strained closed geometry induces a strong electronic deformation within the individual phenyl rings, resulting in quinoidal character (antibonding interactions *within* the phenyl ring and double-bond character *connecting* adjacent phenyl rings). Since the quinoid form is energetically less stable than the aromatic form, the highest-occupied molecular orbital in the smaller nanorings becomes destabilized (i.e., raised in energy), resulting in the optical excitation energy growing larger as a function of nanoring size.

| number of rings [$n$] | [$n$]CPP ($m=0$) | | [$n$]CPPN ($m=1$) | | [$n$]CPPA ($m=2$) | | [$n$]CPPT ($m=3$) | |
|---|---|---|---|---|---|---|---|---|
| | $E_{opt}$ (eV) | oscillator strength | $E_{opt}$ (eV) | oscillator strength | $E_{opt}$ (eV) | oscillator strength | $E_{opt}$ (eV) | oscillator strength |
| 6  | 2.51 | 0.000 | 2.92 | 0.046 | 2.52 | 0.066 | 2.27 | 0.090 |
| 7  | 2.61 | 0.009 | 3.03 | 0.021 | 2.69 | 0.084 | 2.28 | 0.091 |
| 8  | 2.86 | 0.000 | 3.02 | 0.068 | 2.73 | 0.141 | 2.29 | 0.105 |
| 9  | 2.90 | 0.017 | 3.10 | 0.034 | 2.79 | 0.170 | 2.29 | 0.103 |
| 10 | 3.05 | 0.000 | 3.09 | 0.097 | 2.80 | 0.241 | 2.30 | 0.118 |
| 11 | 3.05 | 0.024 | 3.15 | 0.047 | 2.84 | 0.259 | 2.31 | 0.113 |
| 12 | 3.14 | 0.000 | 3.15 | 0.129 | 2.81 | 0.342 | 2.29 | 0.128 |
| 13 | 3.14 | 0.030 | 3.18 | 0.065 | 2.86 | 0.344 | 2.31 | 0.122 |

**Table 1.** Excitation Energies and Oscillator Strengths for the CPP, CPPN, CPPA, and CPPT Nanorings.

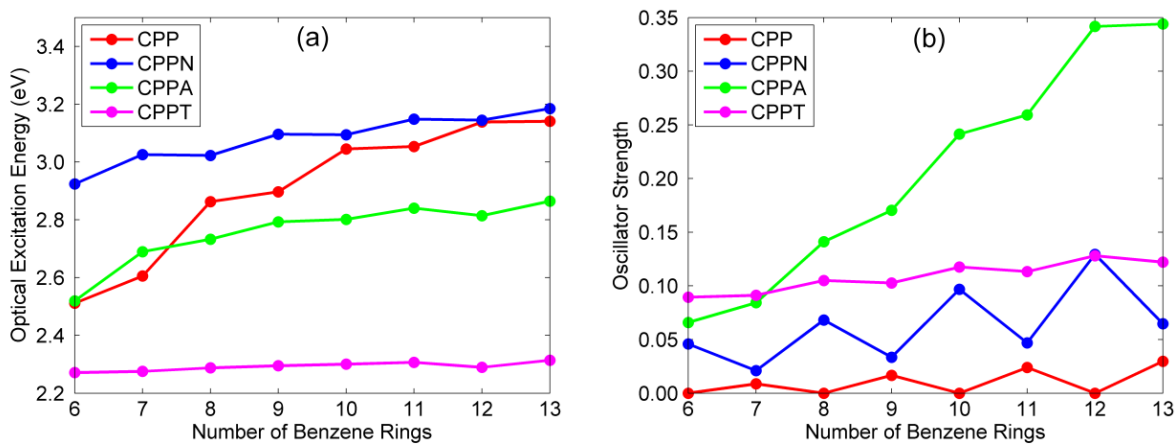

**Figure 2.** (a) Optical excitation energies and (b) oscillator strengths for the various carbon nanorings as a function of size. While all of the nanorings have excitation energies which increase as a function of size in Figure 2a, only the CPPA nanoring has a large, optically-detectable oscillator strength, as shown in Figure 2b.

Although these unusual optical excitations show a clear dependence on molecular size, their spectroscopic observation is, however, determined by the magnitude of the induced oscillator strength. While the CPP nanoring shows the largest variation in excitation energy as a function of size, Figure 2b indicates that these CPP excitations are actually forbidden or optically dark (even-membered CPP rings have zero oscillator strength due to molecular symmetry while odd-membered CPP rings have small but finite oscillator strengths due to reduced symmetry). As depicted in Figure 2b, among all of the nanorings, only the CPPA chiral nanoring has a surprisingly large and distinct oscillator strength which rapidly increases with size.

To provide further insight into these interesting trends, we calculated real-space transition densities using the TD-DFT excited-state density-matrix for each of the $n = 12$ nanorings. In Figure 3, the transition densities give a panoramic view of the orientation and strengths of the individual transition dipole moments throughout the optically-excited structure. For the lowest excitation energy, the transition moments in CPP are aligned sequentially in a head-to-tail arrangement, as shown in Figure 3a. However, because of their circular symmetry, the transition dipole moments are equal in magnitude and effectively cancel to give a net vectorial sum of zero. In the CPPN chiral ring shown in Figure 3b, the transition densities are also delocalized around the phenyl rings, but the very small transition dipole moment on the acene subunit is now oriented along the short axis of the naphthalene (pointing approximately perpendicular to the plane of the nanoring). As a result, the vectorial sum of the CPPN transition moments is now nonzero but still very small – most of the transition moments on the phenyl backbone still nearly cancel each other, as shown in Figure 3b. It is important to mention at this point that all of the acene subunits have transition moments oriented along their short axis with oscillator strengths decreasing in size (0.060, 0.059, and 0.050 for naphthalene, anthracene, and tetracene, respectively). In contrast, the transition densities in the CPPA nanoring are no longer fully delocalized around the entire structure; rather, most of the transition density is concentrated within a semi-circle which includes the anthracene subunit. Due to this particular arrangement of transition moments, the vertical vectorial components in Figure 3c cancel, and only a large horizontal vectorial sum remains. Finally, for the CPPT chiral ring in Figure 3d, the transition density is almost entirely localized to the tetracene unit, and since the oscillator strength of tetracene is minor, the total transition dipole moment of the CPPT nanoring is resultingly small.

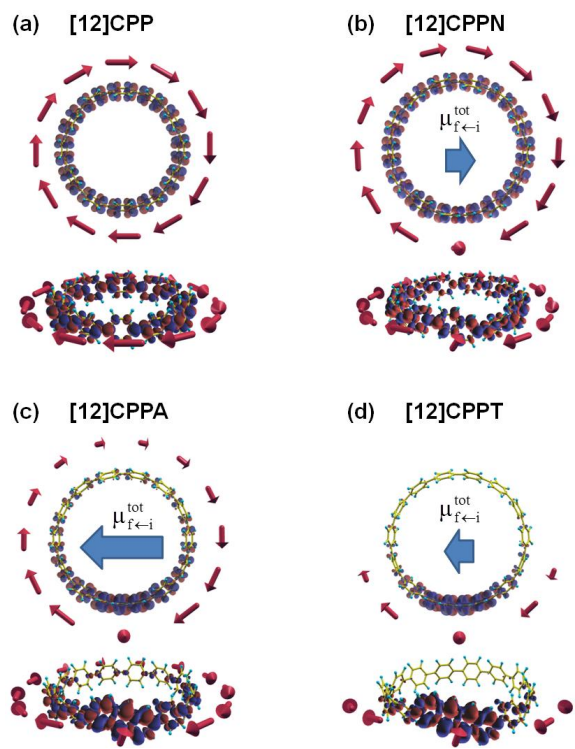

**Figure 3.** Top-down and side views of the TD-DFT transition densities for the (a) [12]CPP, (b) [12]CPPN, (c)[12]CPPA, and (d) [12]CPPT nanorings. The individual transition dipole moments are displayed as red arrows, and their total vectorial sum ($\mu_{f \leftarrow i}^{tot}$) is shown as the blue arrow in the center of each nanoring. Among the various nanorings, only the CPPA ring has a large, horizontal transition dipole.

A question now naturally arises: Why do the transition densities behave so differently in the acene-substituted chiral nanorings (and can we explain these trends in a simple way without resorting to a full TD-DFT transition density analysis)? To address this question, we examine the relative alignment between the highest-occupied and lowest-unoccupied molecular orbital (HOMO/LUMO) energies between the various acenes and the phenylene backbone (it should be noted that similar analyses of HOMO/LUMO alignments are also commonly utilized in the photovoltaics community to rationalize the optical properties of OLED heterojunction systems[1]). In Figure 4, we plot the *non-interacting* HOMO and LUMO energies (relative to vacuum) for the isolated acene molecules and for an isolated phenylene polymer extrapolated to the infinite chain limit. In each of these systems, upon photo-excitation, an exciton is formed in which an electron is promoted into the LUMO, leaving behind a hole in the HOMO. Almost instantaneously, charge migration occurs as the electron naturally jumps from the LUMO of the donor (the material with the higher LUMO energy) to the lower-lying LUMO of the acceptor. Even in this simple zeroth-order picture shown in Figure 4 (the energies depicted here are computed from isolated, non-interacting moieties), we can begin to see why the transition densities in the chiral nanorings are so distinct. For the naphthalene-phenylene heterojuction, which corresponds to the CPPN ring, the naphthalene LUMO lies significantly higher than the phenylene LUMO (0.6 eV), and the excited electron will naturally delocalize into the lower-energy phenylene backbone. In stark contrast, for the tetracene-phenylene heterojunction, the roles of the donor and acceptor are now completely reversed, with the phenylene LUMO lying significantly higher than the tetracene LUMO (0.5 eV). Because of this complete reversal of energetic alignments, the excited electron in the CPPT nanoring will naturally reside primarily on the tetracene unit (which inherently has a small oscillator strength). Finally, the anthracene-phenylene heterojunction presents an extremely unique situation where the HOMO-LUMO gaps are nearly coincident, and the LUMOs of both moieties are energetically aligned. As a result of this alignment, the photo-excited electron will be shared between the anthracene and phenylene subunits, resulting in the transition densities depicted in Figure 3c.

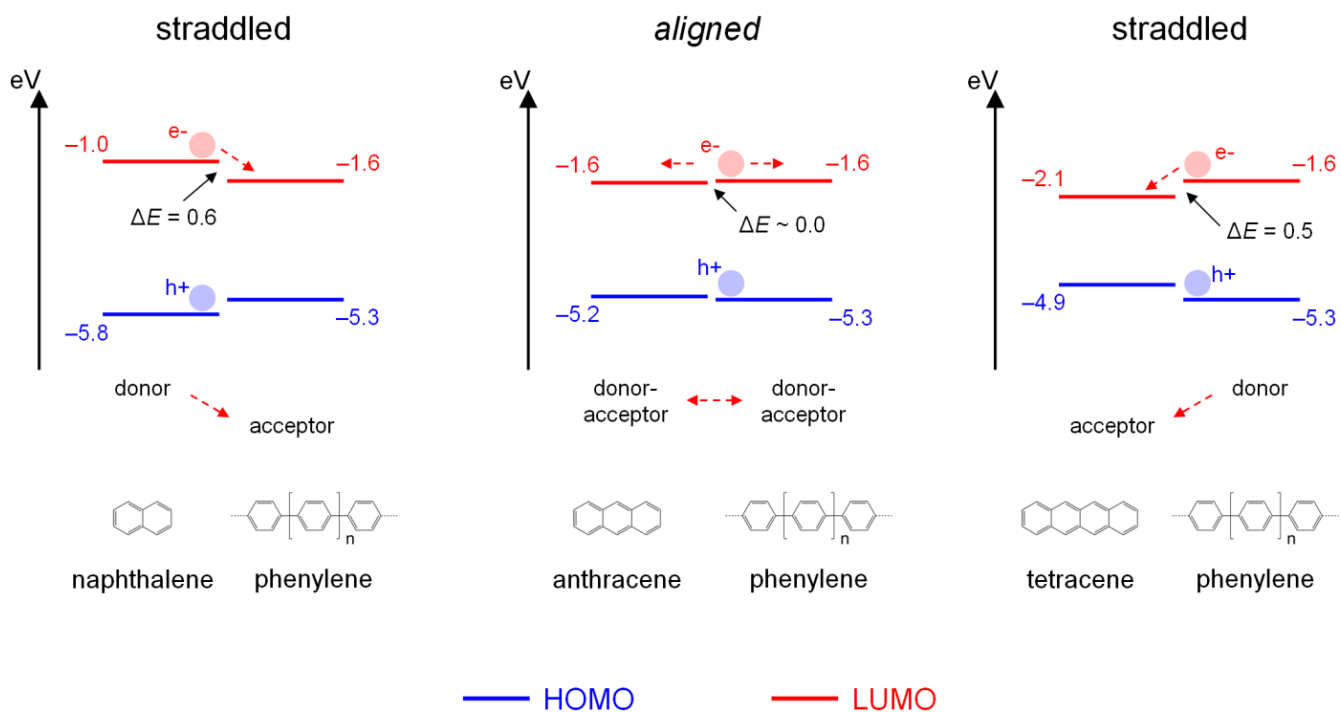

**Figure 4.** HOMO-LUMO energy alignments for each of the heterojunctions found in the various chiral nanorings. Upon photo-excitation, an exciton (electron-hole pair) is created in which the excited electron migrates from the donor LUMO to the acceptor LUMO. Among the various heterojunctions, only the anthracene-phenylene combination has orbital energies that are nearly coincident.

It is important to point out that while other chiral nanorings can also be formed with larger acenes such as pentacene and hexacene, these nanostructures are expected to have even smaller oscillator strengths. Specifically, the LUMOs of the acenes decrease rapidly as a function of size, and based on energetic alignments, the excited electron will be even more localized on the acene subunit which inherently has a small oscillator strength (our benchmark TD-DFT calculations confirm this behavior for nanorings with larger acenes, although we note that extremely long acenes can exhibit multireference/diradical character[20]). Lastly, it is important to mention that we were able to reproduce the same trends in oscillator strengths using different functionals, such as new range-separated DFT methods[16-18]. While the *absolute* HOMO/LUMO energies with respect to vacuum can vary with the choice of functional,[21] we find that the same *relative* energetic alignments between the acene and phenylene unit is still maintained with these other methods (i.e., the CPPA nanoring still has the largest oscillator strength even with different functionals), indicating the large and unique oscillator strengths of the CPPA ring are rather robust and insensitive to the specific functional used.

In conclusion, we have characterized the unusual optoelectronic properties in a series of chiral carbon nanorings which form the fundamental annular segments of carbon nanotubes. Most importantly, we find that *only one ring*, specifically the CPPA nanoring (which corresponds to a ($n+3,n+1$) SWNT segment), is particularly unique and special (i.e., a "ring to rule them all"[23]). Among all of the possible carbon nanorings, only the CPPA ring demonstrates both excitation energies which surprisingly grow larger as a function of size as well as large oscillator strengths which can be readily detected. These unusual optoelectronic properties in CPPA result from a geometric symmetry-breaking combined with an optimal band-gap alignment between the anthracene and phenylene backbone – a unique property which is not present in the other nanorings. As a result, our calculations have several implications for experiments: while the anomalous optoelectronic trends in *symmetric* carbon nanorings (such as CPP) would require low-temperature spectroscopic studies with an applied magnetic field to "brighten" the dark states (via the Aharonov-Bohm effect[22]), the use of these elaborate experimental conditions is not necessary for the *asymmetric* CPPA nanoring system. Specifically, the unusual optoelectronic properties in the CPPA nanoring are *intrinsic* to its geometry and electronic structure, negating the need for large external magnetic fields. As a result, we anticipate that the unique spectroscopic effects in these chiral systems will provide further insight into the subtle interplay between quantum confinement and asymmetry in low-dimensional nanostructures of increasing complexity.

**Acknowledgement.** This research was supported in part by the National Science Foundation through TeraGrid resources (Grant No. TG-CHE1000066N) provided by the National Center for Supercomputing Applications. Funding for this effort was provided by the Laboratory Directed Research and Development (LDRD) program at Sandia National Laboratories, a multiprogram

**Supporting Information Available:** Strain energies, total electronic energies, total enthalpies, and Cartesian coordinates for all of the optimized structures. This material is available free of charge via the Internet at http://pubs.acs.org.